\documentclass[twocolumn]{webofc}
\usepackage[varg]{txfonts}   
\usepackage{graphicx}
\usepackage{amsfonts}
\usepackage{color}
\usepackage{dcolumn}
\usepackage{bm}
\usepackage{braket}
\usepackage[normalem]{ulem}  

\begin{document}
\title{Spontaneous fission half-life in Fm isotopes with nuclear energy density functional}
%
%

\author{\firstname{Kouhei} \lastname{Washiyama}\inst{1}
\fnsep\thanks{\email{washiyama@nucl.ph.tsukuba.ac.jp}} 
}

\institute{Center for Computational Sciences, University of Tsukuba, Tsukuba, 305-8577, Japan
}

\abstract{%
A microscopic description of fission dynamics is important to understand the decay properties of neutron-rich heavy nuclei
that are relevant to $r$-process nucleosynthesis.
To provide a reliable and efficient method to evaluate the spontaneous fission half-life, 
we develop a method, called the constrained Hartree--Fock--Bogoliubov (CHFB) plus local quasiparticle random-phase approximation (LQRPA), to include dynamical residual effects in the collective inertia.
With the CHFB + LQRPA, we evaluate the collective potential and the collective inertia along a mass-symmetric fission path in Fm isotopes with the neutron numbers $N=158$--164.
The obtained LQRPA inertia is much larger than the cranking one 
that ignores dynamical residual effects
and shows a remarkable variation along the fission path. 
We estimate the fission half-life of the Fm isotopes 
using the action integral with the obtained collective potential and inertia.
A large difference between the fission half-lives obtained with the LQRPA inertia and with the cranking inertia is observed.
This indicates the importance of evaluating the collective inertia
for estimating the fission half-life.
}
\maketitle

\section{Introduction}
\label{intro}

Nuclear fission attracts much interest
in various subjects \cite{andreyev17}.
One of the recent applications on fission
is the nuclear reaction network calculations of the $r$-process nucleosynthesis simulations
\cite{mumpower16,giuliani18a}
since low-energy fission in the $r$-process environment
may determine the end point of the $r$-process nucleosynthesis
and a part of the abundance patterns of the elements in the universe.
The network calculations involve experimentally unknown nuclei.
Therefore, a microscopic model for fission
is necessary to describe fission dynamics 
of both stable and unstable nuclei \cite{bender20}.

The nuclear energy density functional (EDF) theory has been 
widely used for various objectives in nuclear physics
as a microscopic model.
The nuclear EDF gives a good description of the ground-state property of nuclei in the whole nuclear chart \cite{bender03,nakatsukasa16}.
Recently, a microscopic description for fission
has been developed with EDF approaches
\cite{warda02,bonneau04,staszczak09,abusara10,baran11,staszczak13,sadhukhan13,rodriguez-guzman14}.
In these EDF approaches,
spontaneous or low-energy fission has been described 
with the Wenzel--Kramers--Brillouin (WKB) approximation
for quantum many-body tunneling.
The potential energy and the collective inertia used in the WKB approximation have been calculated with the constrained Hartree--Fock--Bogoliubov (HFB) method with constraint on the collective variables
and with the so-called cranking approximation \cite{girod79,ring-schuck}, respectively.
It is well known that
the cranking approximation ignores dynamical residual effects,
particularly time-odd terms of the EDF,
in the collective inertia.
This gives a significant deviation from
the collective inertia that includes the dynamical residual effects \cite{dobaczewski81}.
However, the cranking approximation has been widely used 
because of its low computational cost.


In a microscopic derivation of the five-dimensional quadrupole collective Hamiltonian \cite{bohr},
the constrained HFB (CHFB) plus local quasiparticle random-phase approximation (LQRPA) \cite{hinohara10} 
was proposed to properly include the dynamical residual effects in the inertial functions.
The CHFB + LQRPA has been applied 
to the construction of the collective Hamiltonian in triaxial shapes 
with the semi-realistic pairing-plus-quadrupole Hamiltonian 
\cite{hinohara10,hinohara11,sato11,hinohara12,sato12}.
To limit axially symmetric shapes,
the collective Hamiltonian was constructed 
with the Skyrme EDF \cite{yoshida11}.

Recently, we applied the CHFB + LQRPA method with the Skyrme EDF
to the description of the collective inertia
along the mass-symmetric fission path and showed the importance of the improved collective inertia in the description of the fission dynamics \cite{washiyama21}.
To overcome a problem of high computational cost 
to evaluate the collective inertia with the LQRPA, 
we employed the finite amplitude method (FAM) \cite{nakatsukasa07,inakura09,avogadro11}. 
In this paper, we extend this work
to the description of the fission half-life in Fm isotopes.

\section{Method}

Since the formulation of the CHFB + LQRPA for the collective inertia along the fission path was presented in Refs. \cite{hinohara10,washiyama21},
we give in this section a brief explanation of the formulation.
Here, we assume one collective coordinate $q$
to describe the fission dynamics.

The collective potential is obtained 
by solving the CHFB equation with a given constraining operator $\hat{s}$,
\begin{align}
  \delta \langle \phi(s)| \hat{H}_{\text{CHFB}} |\phi(s) \rangle &= 0, \label{eq:CHFB} \\ 
  \hat{H}_{\text{CHFB}} &= \hat{H}_{\text{HFB}} - \sum_{\tau=n,p} \lambda_\tau \hat{N}_\tau - \lambda_s \hat{s} ,
\end{align}
where $\lambda_{n,p}$ and $\hat{N}_{n,p}$ denote the Fermi energies 
and the number operators, respectively, for neutrons and protons,
and $\lambda_s$ is a Lagrange multiplier
of constraining the collective variable $s=\braket{\phi(s)|\hat{s}|\phi(s)}$.
The energy minimization in Eq.~(\ref{eq:CHFB})
leads to the CHFB state $|\phi(s)\rangle$
and collective potential $V(s)=\langle \phi(s)|\hat{H}_{\text{HFB}}|\phi(s)\rangle$.

The collective inertia is determined by local normal modes build on CHFB states.
On top of the CHFB state $|\phi(s)\rangle$,
the LQRPA equations, 
\begin{align}
	\delta \langle \phi(s)| [\hat{H}_{\text{HFB}}, \hat{Q}_i(s)]
	- \frac{1}{i}\hat{P}_i(s) |\phi(s)\rangle &= 0, \\
	\delta \langle \phi(s)| [\hat{H}_{\text{HFB}}, \frac{1}{i}\hat{P}_i(s)] -
	\Omega_i^2(s)\hat{Q}_i(s) |\phi(s)\rangle &= 0,
\end{align}
are solved 
to determine the local generators $\hat{Q}_i(s)$ and $\hat{P}_i(s)$ defined at $s$.  
Here, $\Omega_i^2(s)$ is the squared eigen-frequency of the local normal mode.
Once the local normal mode is selected, 
the collective kinetic energy is expressed in terms of the collective variable $s$ as \cite{hinohara10} 
\begin{align}
T= \frac{1}{2} \dot{q}^2 = \frac{1}{2} \mathcal{M}(s) \dot{s}^2, 
\end{align}
where the collective inertia $\mathcal{M}(s)$ 
is defined as
\begin{equation}\label{eq:vibrational-mass}
\mathcal{M}(s) =
\frac{d q}{d s} \frac{d q}{d s}.
\end{equation}
The inverse of the derivative 
is evaluated using the local generator $\hat{P}_i(s)$ as
%
\begin{align}\label{eq:derivative}
	\frac{d s}{d q} = \frac{d }{d q}
   \langle \phi(s) | \hat{s}|\phi(s)\rangle 
   =  \langle \phi(s) | [\hat{s}, \frac{1}{i}\hat{P}_{i}(s)]|\phi(s)\rangle.
\end{align}
Thus, the collective inertia \eqref{eq:vibrational-mass}
is determined by the local normal mode of the LQRPA equations.
Note that the above formulation can be generalized to 
the cases with more than one collective coordinates \cite{hinohara10}.

As a constrained operator
to describe the coordinate of the mass-symmetric fission path,
the axially symmetric mass quadrupole operator, 
\begin{equation}
   \hat{s}=\hat{Q}_{20} = \sqrt{\frac{16\pi}{5}}\sum_{i=1}^A r_i^2 Y_{20}(\hat{\boldsymbol{r}}_i) ,
\end{equation}
is used. 
To limit axially symmetric shapes, the triaxiality $\braket{\hat{Q}_{22}} = 0$ is kept, which is defined as
$\hat{Q}_{22}=\sqrt{16\pi/5}\sum_{i=1}^A r^2_i [Y_{22}(\hat{r}_i) +Y_{2-2}(\hat{r}_i)]/\sqrt{2}$.
For the potential energy,
we solve the CHFB equation with the two-basis method \cite{gall94,terasaki95}
in a three-dimensional Cartesian mesh.
%
Enforcing the reflection symmetries 
about the $x=0$, $y=0$, and $z=0$ planes,
we use a numerical box of 11.5\,fm${}\times 11.5$\,fm$ {}\times 16.5$\,fm 
in $x > 0$, $y > 0$, $z > 0$ 
with a mesh size of 1.0\,fm.
With these reflection symmetries, odd multipole moments (e.g., octupole moments) can not be considered.
Namely, mass-asymmetric fission paths are not in the present scope.
The single-particle basis consists of
1820 neutron and 1440 proton Hartree-Fock basis states 
to achieve the maximum quasiparticle energy $E_{\text{QP}}^{\text{max}}\approx 60$ MeV.
For the CHFB + LQRPA calculations,
we follow Ref. \cite{washiyama21}
and use the FAM \cite{nakatsukasa07,inakura09,avogadro11},
in particular the FAM with contour integration technique developed in Ref. \cite{hinohara13}, 
for the solution of the LQRPA equations.
We choose the most collective mode among LQRPA modes in low frequencies $\Omega^2_i < 16$\,MeV$^2$
following the prescription in Ref. \cite{washiyama21}.
We solve the CHFB + LQRPA in 
$Q_{20}=\braket{\hat{Q}_{20}} \le 110$\,b with $\Delta Q_{20}=5$\,b.
We used the SkM$^*$ EDF \cite{bartel82} and 
the contact volume pairing with a pairing window of 20\,MeV described in Ref. \cite{ev8new}.
The pairing strengths were adjusted separately for neutrons and protons to reproduce the empirical pairing gaps in $^{256}$Fm.

\section{Results and Discussions}
\label{sec:result}

We show the results 
of the mass-symmetric fission for the even--even Fm isotopes 
with the neutron numbers $N=158$--164.
The reason why we chose $^{258,260,262,264}$Fm is that 
the mass distributions of fission fragments in $^{258,259}$Fm are symmetric
and experimental systematic observation in the mass region around Fm isotopes suggests the mass-symmetric fission in Fm isotopes with $N \ge 158$ \cite{lane96}.
We did not investigate odd-$N$ Fm isotopes 
because of difficulty to handle odd nuclei
in the present EDF framework.

\subsection{Collective potential and pairing gaps}
\label{sec:result-barrier}

\begin{figure}[tb]
\centering
\includegraphics[width=0.95\linewidth,clip]{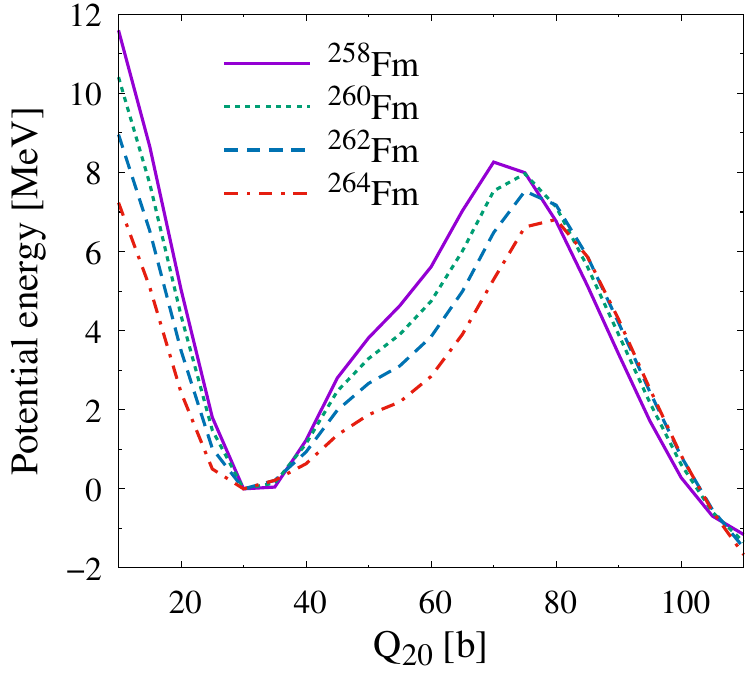}
\caption{
Collective potential energy 
as a function of $Q_{20}$ for Fm isotopes
calculated with the CHFB.
The solid, dotted, dashed, and dot-dashed lines denote the results of $^{258}$Fm, $^{260}$Fm, $^{262}$Fm, and $^{264}$Fm, respectively.
The collective potential energy is shifted so as to make its minimum energy zero.
}
\label{fig:fissionbarrierFm}
\end{figure}

\begin{figure}[tbh]
\centering
\includegraphics[width=0.9\linewidth,clip]{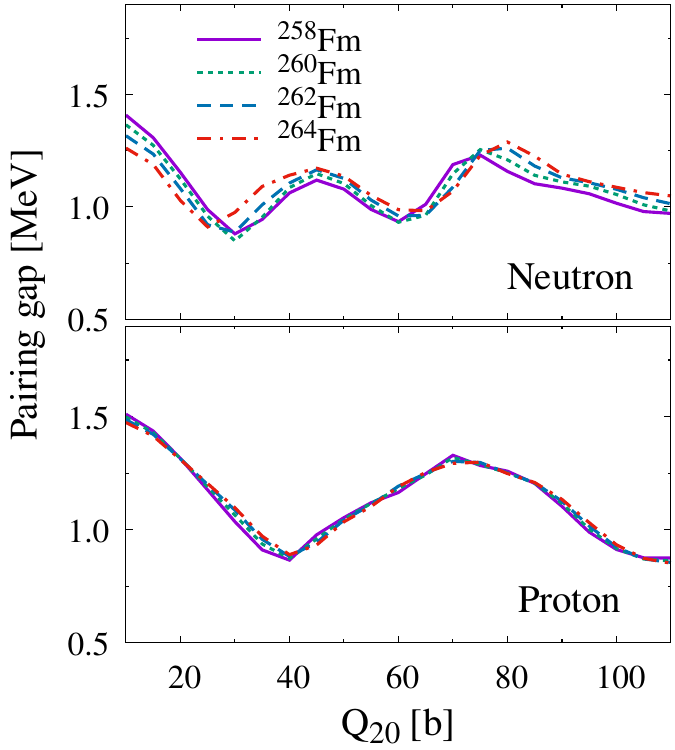}
\caption{
Neutron (top) and proton (bottom) 
pairing gaps as a function of $Q_{20}$ for 
the Fm isotopes.
}
\label{fig:pairinggapFm}
\end{figure}

First, we show the collective potentials of the Fm isotopes
calculated with the CHFB method
along the mass-symmetric fission path.
Figure~\ref{fig:fissionbarrierFm} shows
the collective potential 
as a function of $Q_{20}$ for $^{258,260,262,264}$Fm.
The obtained potential energies are shifted 
so as to make the minimum energy
of each nucleus set to be zero.
The collective potential shows a neutron-number 
dependence and as the neutron number increases,
the fission barrier height decreases
and the value of $Q_{20}$ at the barrier position increases.
The value of $Q_{20}$ at the minimum energy
is similar to each other.
In each nucleus,
the energy becomes negative at $Q_{20} \ge 105$\,b.

Figure~\ref{fig:pairinggapFm}
shows the neutron and proton pairing gaps in the Fm isotopes.
In neutron, the variation of the gap on $Q_{20}$ is similar to each other.
In the proton gaps, almost the same $Q_{20}$ dependence is obtained among different Fm isotopes.

\subsection{Collective inertia}

\begin{figure}[tbh]
\centering
\includegraphics[width=0.95\linewidth,clip]{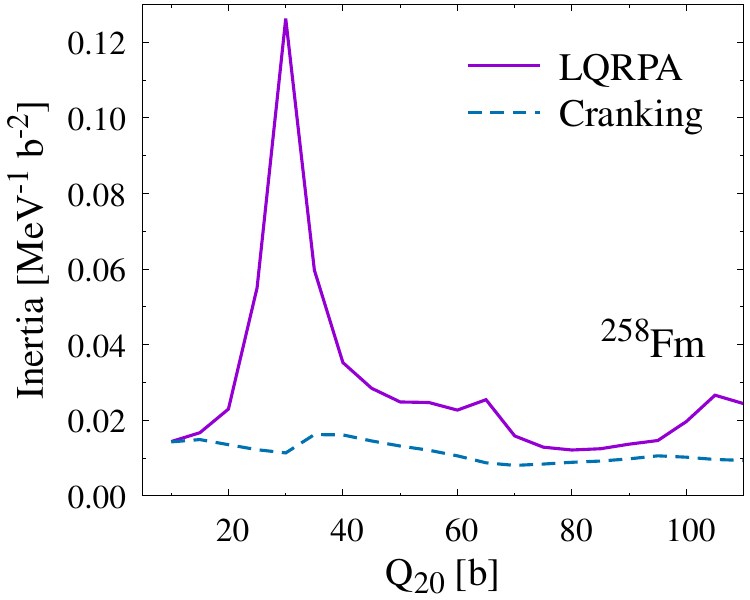}
\caption{
Collective inertia as a function of $Q_{20}$ for $^{258}$Fm.
The solid and dashed lines correspond to the collective inertia
obtained with the LQRPA and 
the one with the perturbative cranking approximation, respectively.
}
\label{fig:inertia_QRPA_PC_Fm258} 
\end{figure}

Next, we discuss the collective inertia along the fission path
in the Fm isotopes.
Figure~\ref{fig:inertia_QRPA_PC_Fm258} shows the collective inertia
obtained with the CHFB + LQRPA method
denoted as the solid line
as a function of $Q_{20}$ for $^{258}$Fm.
A sharp increase at $Q_{20} \approx 30$ b, where the ground state is located, is seen.
A similar structure to this was presented in $^{240}$Pu and $^{256}$Fm
in Ref. \cite{washiyama21}.
In Fig.~\ref{fig:inertia_QRPA_PC_Fm258}, the collective inertia with the perturbative cranking approximation that ignores the dynamical residual effects
is shown by the dashed line.
At all $Q_{20}$, 
the LQRPA inertia is larger than the cranking one,
and a variation of the LQRPA inertia is remarkable.
The dynamical residual effects significantly increase the collective inertia.

\begin{figure}[tb]
\centering
\includegraphics[width=0.95\linewidth,clip]{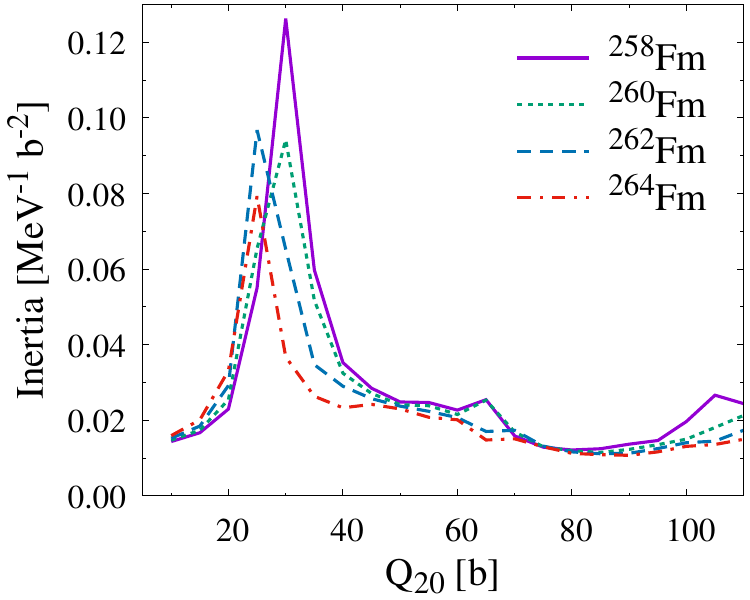}
\caption{
Collective inertia obtained with the CHFB + LQRPA method 
for different Fm isotopes.}
\label{fig:inertia_Fmisotopes} 
\end{figure}

Figure \ref{fig:inertia_Fmisotopes} shows the LQRPA collective inertia
for the Fm isotopes.
The collective inertia for all the isotopes has similar $Q_{20}$ dependence:
a sharp increase at $Q_{20}\approx 30$\,b
and smooth behavior around $Q_{20} \approx 80$\,b, 
at the top of the fission barrier.

\subsection{Spontaneous fission half-life}

From the collective potential and the collective inertia along the fission path,
the spontaneous fission half-life can be estimated.
The fission half-life is given by
\begin{equation}
T_{1/2} = \ln 2/(nP),
\label{eq:halflife}
\end{equation}
where $n$ is the number of assaults on the fission barrier per unit time.
According to the former works \cite{sadhukhan13},
$n=10^{20.38}$\,s$^{-1}$ is employed.
$P$ denotes the penetrability given as
\begin{equation}
  \quad P = [1 + \exp(2S)]^{-1},
  \label{eq:probability}
\end{equation}
where the action $S$ is given in the WKB approximation as
\begin{equation}
  S = \int_{Q_{\text{in}}}^{Q_{\text{out}}} d Q_{20}
      \sqrt{2\mathcal{M}(Q_{20})[V(Q_{20})-E_0]},\label{eq:action}
\end{equation}
where $Q_{\text{in}}$ and $Q_{\text{out}}$ are the classical inner and outer turning
points along the fission path, respectively,
and $E_0$ denotes the ground-state energy.

\begin{figure}[htb]
\centering
\includegraphics[width=\linewidth,clip]{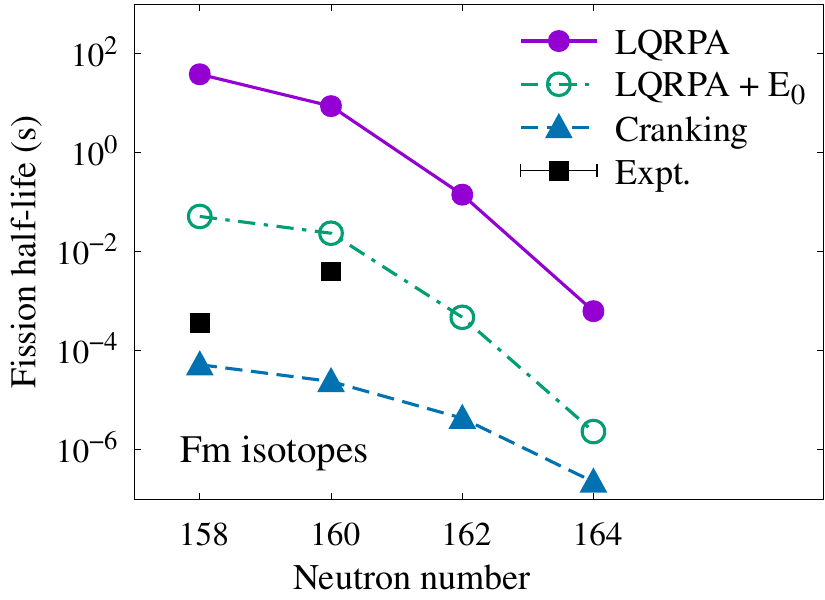}
\caption{
Spontaneous fission half-life obtained with the LQRPA inertia by the filled-circle solid line
and with the cranking inertia by the filled-triangle dashed line
for the Fm isotopes, compared with the experimental data \cite{fission-halflife00} by the filled squares.
The open-circle dot-dashed line shows the result of the LQRPA inertia
by adding a zero-point energy correction of $E_0=0.5$\,MeV (see text).
}
\label{fig:fission_halflife_Fm}
\end{figure}

Figure \ref{fig:fission_halflife_Fm} shows the spontaneous fission half-life
obtained with the LQRPA inertia (filled-circle solid line) and
with the cranking inertia (filled-triangle dashed line)
in the Fm isotopes.
In these results, the ground-state energy is set $E_0=0$.
The available experimental data are plotted by squares \cite{fission-halflife00}.
Note that the experimental data point for $^{260}$Fm may be questionable,
see the NUBASE2020 evaluation \cite{Nubase2020}
for detail.
%
%
The fission half-life obtained with the LQRPA inertia is 4--6 orders of magnitude larger than that with the cranking inertia.
This difference originates from only the difference in the collective inertia with or without the dynamical residual effects.
The fission half-life is significantly sensitive to the choice of the collective inertia.
The fission half-lives obtained with both the LQRPA inertia and cranking one decrease as the neutron number increases,
which is expected from the collective potential and
the collective inertia.
The spontaneous fission half-life with the LQRPA inertia deviates from the experimental one by several orders of magnitude.

\subsection{Discussions}
The significant deviation between the fission half-life
in the theory and experimental one
indicates a room for the improvement of the theory.
To improve the description of the fission half-life,
the following points can be considered.
%
%

First point is to consider zero-point energy corrections
to the ground-state energy $E_0$
and/or to the collective potential energy
in the action in Eq.~\eqref{eq:action}.
In some of former works \cite{staszczak09,staszczak13,sadhukhan13,rodriguez-guzman14},
a constant value has been employed for the ground-state energy $E_0$
and/or the vibrational and rotational zero-point energy corrections
to the collective potential energy
have been considered
by using the Gaussian overlap approximation 
to the generator-coordinate method.
%
In this paper, we tested a constant value for the ground-state energy of $E_0=0.5$\,MeV.
The result is shown by the open-circle dot-dashed line in Fig.~\ref{fig:fission_halflife_Fm}, denoted as LQRPA + $E_0$.
This simple correction in the ground-state energy decreases
the fission half-life in about 3 orders of magnitude.
It is however noticed that
the description of the zero-point energy correction
should be consistent with the estimation of 
the number of assaults $n$ 
in Eq.~\eqref{eq:halflife}
and further considerations are necessary.

\begin{figure}[tb]
\centering
\includegraphics[width=\linewidth,clip]{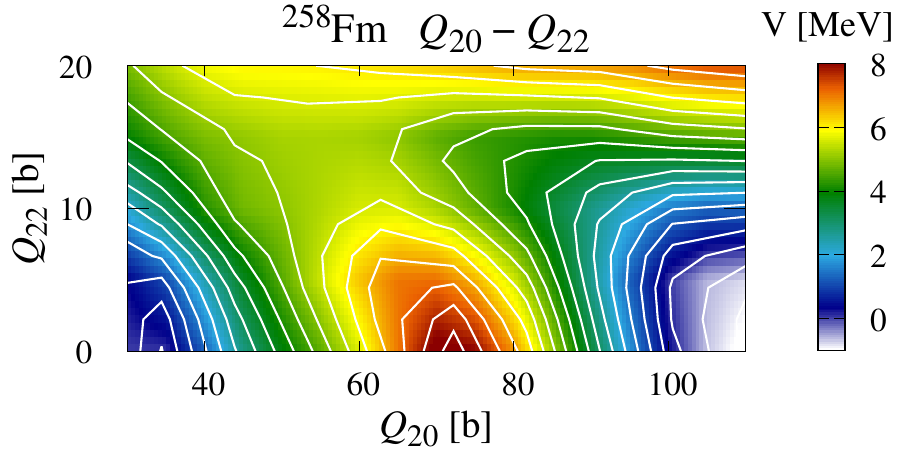}
\caption{
Potential energy surface in the $Q_{20}$--$Q_{22}$ plane
for $^{258}$Fm.
The minimum energy is set to be zero.
The contour interval is 0.5\,MeV.
}
\label{fig:Q20Q22Fm258PES}       
\end{figure}

Second point
is to take more than one collective variable into account
for fission dynamics.
Former works showed that
inclusion of triaxial shapes 
reduces the fission barrier height by a few MeV
in Fm isotopes \cite{warda02,staszczak09,abusara10}.
This reduction is expected to give a decrease in the action $S$,
and decrease in the fission half-life.
Figure \ref{fig:Q20Q22Fm258PES} shows the potential energy surface as functions of $Q_{20}$ and triaxiality $Q_{22}$
calculated with the CHFB with the SkM$^*$ EDF for $^{258}$Fm.
One clearly sees that
the height of the fission barrier 
is reduced by about 3\,MeV
by including triaxial deformations with $Q_{22} > 0$.
%
Then, the fission path that corresponds to the minimized potential
changes from the axially symmetric shape to triaxial shapes.
It is noticed that 
since the collective inertia obtained with the CHFB + LQRPA
much more strongly depends on the collective variables than the cranking inertia does,
the least-action path in Eq.~\eqref{eq:action}
might be different from the path 
that corresponds to the minimized potential.
This will affect the description of the fission half-life.

\section{Summary}


In this paper, we investigated the fission half-lives in
the even--even Fm isotopes with $N=158$--164,
in which the mass-symmetric fission mode is expected to be dominant.
We performed the CHFB + LQRPA calculations to obtain
the collective potential and inertia along the fission path
and perform the action integral to obtain the fission half-life.
We clearly show that the improved description of the collective inertia changes the story of spontaneous fission.
Namely, compared with the result obtained with the cranking inertia,
the LQRPA inertia becomes significantly larger, and then 
makes the fission half-life longer.
However, a large discrepancy between the fission half-life with the CHFB + LQRPA and
experimental one is found.
Including the zero-point energy correction
and other collective variables than the axially symmetric quadrupole moment $Q_{20}$
might give a better description of the fission half-life.
The work along these lines is in progress.

\section*{Acknowledgments}
Numerical calculations were performed using computational resources of Wisteria/BDEC-01 Odyssey (the University of Tokyo), provided by the Multidisciplinary Cooperative Research Program in the Center for Computational Sciences, University of Tsukuba.

%
%
%
%
%

\end{document}